\documentclass[
    ,final            % use final for the camera ready runs
  ]
  {aipproc}

\layoutstyle{6x9}

\begin{document}

\title{Effective field theory and electro-weak processes}

\classification{ 23.40.-s 
%<PACS numbers;   \texttt{http://www.aip..org/pacs/index.html}>
}

\author{F. Myhrer}{
  address={Dept. Physics and Astronomy, Univ. South Carolina, Columbia, SC 29208, USA}
}
%
%\author{<author2>}{
%  address={<common address for author2 and author3>}
%}
%
%\author{<author3>}{
%  address={<common address for author2 and author3>}
%  ,altaddress={<author1 address>} % additional visiting address
%}
%
%\today

\begin{abstract}
Heavy baryon chiral perturbation theory is applied to 
one- and two nucleon processes. 
\end{abstract}

\maketitle

%%%%%%%%%%%%%%%%%%%%%%%%%%%%%%%%%%%%%%%%%%%%
%% MAINMATTER
%%%%%%%%%%%%%%%%%%%%%%%%%%%%%%%%%%%%%%%%%%%%

\subsection*{Introduction}

A careful and systematic study of low-energy  
weak- and strong interaction reactions 
is desirable  in order 
to enhance our understanding of some fundamental 
astro-physical processes. Since low 
energy processes are insensitive to details 
of the short distance structures of the hadrons, 
we can make use of an effective field theory like  
Chiral Perturbation Theory (ChPT), which 
allows a unified approach 
to weak- and strong interaction processes. 
The  ChPT Lagrangian,  which 
reflects the symmetries and  the 
symmetry breaking pattern 
of the underlying theory of QCD, 
also gives a model-independent, 
gauge-invariant evaluation of radiative QED corrections to these 
reactions. 

We know that 
the QCD lagrangian is chirally symmetric
provided 
the $u$ and $d$ quarks are massless. 
Furthermore, it is established that chiral symmetry is spontaneously broken, 
which implies the existence of 
massless Goldstone Bosons (pions). 
The quarks have non-zero masses which however are small  
compared to the QCD scale,
 $m_u \simeq m_d \ll \Lambda_{QCD}$.   
Therefore, a perturbative treatment of the explicit chiral 
symmetry breaking appears reasonable. 
In ChPT these considerations are reflected in the 
hadronic scale,  $\Lambda_{ch} \simeq  1$ GeV $\simeq m_N$, 
being much larger than   
the corresponding pion mass 
 $m_\pi  \; (\propto \sqrt{m_{quark}} ) \ll \Lambda_{ch}  $. 
ChPT assumes that we consider only low-energy reactions which 
only allow low momentum probes.
As a result we will consider the following 
(perturbative) expansion parameter in 
Heavy Baryon Chiral Perturbation Theory (HBChPT): 
$%\varepsilon = 
Q/\Lambda_{ch} \ll 1$, 
where  $Q$ denotes either the 
%process' 
typical 4-momentum involved in the process under consideration, 
or $m_\pi$.  

%\section{<Another section>} 

\subsection*{The HBChPT Lagrangian}

${\cal L}_{ch}$ is written as an expansion in powers 
of $Q/\Lambda_{ch}$, see e.g. the reviews~\cite{Bernard:1995,Bernard:2008}
\begin{eqnarray}
{\cal L}_{ch} &=& {\cal L}^{(1)}_{\pi N} + 
{\cal L}^{(2)}_{\pi N} + {\cal L}^{(2)}_{\pi \pi} + 
{\cal L}^{(3)}_{\pi N} + \cdots 
\nonumber
\end{eqnarray} 
where ${\cal L}^{(\nu)}$ contains terms of order  
$\left(Q/\Lambda_{ch}\right)^\nu$. 
We assume that the terms in the lowest order Lagrangian  
give  the dominant 
contributions to a process.  
The higher order terms presumably give smaller  
perturbative corrections. In HBChPT the 
pions are treated relativistically, whereas the 
nucleons are treated non-relativistically. 
In reality we have two simultaneous expansion parameters,  
$\left(Q/\Lambda_{ch}\right)^\nu$ and $\left(Q/m_N\right)^\nu$, 
which for pragmatic purposes are considered simultaneously. 

{\it The lowest order pion Lagrangian} is: 
\begin{eqnarray}
{\cal L}^{(2)}_{\pi \pi} &=& 
\frac{f_\pi^2}{4} \; Tr\Big\{ 
\nabla_\mu U^\dagger \nabla^\mu U + 
\chi^\dagger U + \chi U^\dagger 
\Big\} \; , 
\label{eq:pion}
\end{eqnarray} 
where $\nabla_\mu U = \partial_\mu U - i(v_\mu +a_\mu)U +iU(v_\mu-a_\mu)$. 
Here  
 $v_\mu$ and $a_\mu$ are external currents, and 
 $\chi \propto \left(\begin{array}{cc}
 m_u   & 0   \\ 
 0     & m_d \\ 
\end{array} 
\right)  
$.
In the evaluations of specific processes 
the $U$-field is expanded in powers of the pion field:  
$U = uu = exp(i\vec{\tau}\cdot \vec{\phi})/f_\pi 
\simeq 1 + (i\vec{\tau}\cdot \vec{\phi} ) /f_\pi +\cdots$. 
This expansion gives 
the familiar first two terms in ${\cal L}^{(2)}_{\pi \pi}$: 
\begin{eqnarray}
{\cal L}^{(2)}_{\pi \pi}&=& 
\frac{1}{2} \left(\partial_\mu \vec{\phi}\; \right)^2 - 
\frac{1}{2} m_\pi^2 \; \vec{\phi}^{\; 2} + \cdots 
\nonumber
\end{eqnarray} 
 
%Some url test \url{http://www.world.universe}.  

{\it The lowest order heavy nucleon Lagrangian} is:
\begin{eqnarray}
{\cal L}_{\pi N}^{(1)} = 
\bar{N} \Big\{ i\left(\upsilon \cdot {\cal D} \right) +g_A \left( S\cdot u \right)
\Big\} N \; ,
\label{eq:piN}
\end{eqnarray} 
where ${\cal D}^\mu = \partial^\mu + 
\frac{1}{2}[u^\dagger, \partial^\mu u] 
-\frac{i}{2}u^\dagger (v^\mu+a^\mu)u-\frac{i}{2}u(v^\mu-a^\mu)u^\dagger $. 
If we choose the nucleon velocity $\upsilon ^\mu =(1,\vec{0})$, then 
the nucleon spin is: 
$S^\mu = (0,\frac{1}{2}\vec{\sigma})$. 
By again expanding the $U$-field we find the following first three terms: 
\begin{eqnarray}
{\cal L}^{(1)}_{\pi N} &=& 
\bar{N} \left\{ 
i\frac{\partial}{ \partial t } - 
\frac{ \vec{\tau}\cdot \left(\vec{\phi}\times \vec{\dot{\phi}}\; \right) }
{4f_\pi^2} 
+ \frac{g_A}{2f_\pi} \vec{\tau} \cdot 
\left({\bf \sigma}\cdot {\bf \nabla}\vec{\phi}\; \right)
\right\} N + \cdots 
\nonumber 
\end{eqnarray}
In effective field theory the 
lagrangian contains low energy constants (LECs),
which parametrize the short-distance physics not probed at 
long wave-lengths. 
%
%These  LECs must be determined in order for 
%the theory to have predictive power. 
In principle a LEC should be evaluated from QCD but in practice 
LECs are determined by reproducing the experimental values of 
appropriate observables.  
The nucleon axial coupling constant, $g_A \simeq 1.27$, 
in Eq.(\ref{eq:piN}) 
is an example of a LEC.  
Once the LECs are determined the 
theory has predictive power.
%lagrangian can make predictions for other processes.

{\it In the next order heavy nucleon Lagrangian} 
with the expanded $U$-field,  
\begin{eqnarray}
{\cal L}^{(2)}_{\pi N} &=& \bar{N} \left\{ 
\frac{\left(v\cdot \partial \right)^2 -\partial^2} 
{2m_N} + \cdots  
\right\} N \; , 
\nonumber
\end{eqnarray}
we display only the heavy nucleon kinetic operator 
(``the Schr\"odinger kinetic operator") 
%\begin{eqnarray}
$\frac{\vec{\nabla}^{\; 2} }{2 m_N} $.
%\nonumber 
%\end{eqnarray} 
This nucleon kinetic operator is a ``recoil" correction 
to the leading terms; 
in other words the heavy 
nucleon expansion {\it is different} from the 
Foldy-Wouthuysen expansion as discussed in~\cite{Gardestig:2007}.  
In the following we will give some examples of one- and two-nucleon 
electroweak processes 
which have been evaluated in HBChPT. 

%%%%%%%%%%%%%%%%%%%%%%%%%%%%%%%%%%%%%%%%%%%%%%%%%%%%%%%%%%%%%%%%%%%
\subsection{Specific processes}
%\subsubsection*{One-nucleon processes}%%%%%%%%%%%%%%%%%%%%%%%%%%%%
The following {\it one-nucleon processes},  
%\begin{itemize}
%\item
ordinary muon capture:  
$\mu^- + p \to \nu_\mu + n$ (OMC), 
%\item 
radiative muon capture: 
$\mu^- + p \to \nu_\mu + n + \gamma$ (RMC),  
%\item 
and the radiative corrections to 
$n\to p + e^- + \nu_e$ and $\bar{\nu}_e +p \to e^+ + n$ (the CHOOZ process),  
have all been investigated in HBChPT. 
%\end{itemize} 
Since in all these weak-interaction processes the momentum transfers are small 
$Q \ll m_W$,  
the effective interaction lagrangian is the ``Fermi" Lagrangian: 
\begin{eqnarray}
{\cal L}_{Fermi} = \frac{G_F}{\sqrt{2}} \; J_{\beta}(lepton) \cdot 
J^\beta (hadron)
\nonumber
\end{eqnarray} 
where
%\begin{eqnarray}
$
J_\beta (lepton) = \bar{u}_\nu \gamma_\beta (1-\gamma_5) u_l 
%= v_\beta^{lep} - a_\beta^{lep} 
$
%\; \; \; {\rm 
and
%} \; \; \; 
%\nonumber \\  
$
J_\beta (hadron) = v_\beta^{had} - a_\beta^{had}
$. 
%\nonumber
%\end{eqnarray}
Traditionally the hadronic currents $ v_\beta^{had}$ and $ a_\beta^{had}$  
are written as:
\begin{eqnarray}
v_\beta^{had}  &=& \bar{\Psi}\left\{ 
 G_V(q^2)\gamma_\beta +  G_M(q^2)
\frac{i\sigma_{\beta\delta}q^\delta}
{2m_N} + {\rm 2nd \; \; class} 
\right\} \Psi 
\nonumber \\ &&
\nonumber \\ 
a_\beta^{had}  &=& \bar{\Psi}\left\{ 
G_A(q^2)\gamma_\beta \gamma_5 + 
G_P(q^2)\frac{q_\beta\; \gamma_5 }
{2m_N} + {\rm 2nd \; \; class} 
\right\} \Psi \; .  
\nonumber 
\end{eqnarray}
When we expand the nucleon form-factors including the $q^2$ terms,  
the LECs are determined by 
the nucleon's  r.m.s. radius,  axial radius, 
anomalous nucleon magnetic moments, i.e. 
$G_M(q^2) = \kappa_p - \kappa_n $, 
and the Goldberger-Treiman discrepancy.  
The pseudo-scalar form factor is derived in ChPT 
(including one-loop corrections) and found to be 
\begin{eqnarray}
\frac{G_P(q^2)}{2 m_N} &=& 
-\frac{2f_\pi \ g_{\pi NN} }{q^2-m_\pi^2} - \frac{1}{3} 
g_A \ m_N <r_A^2 > \; , 
\label{eq:RMC}
\end{eqnarray} 
where the values of all parameters in Eq.(\ref{eq:RMC}) have been 
determined from other reactions. 
This 
expression for $G_P(q^2)$ was derived some time ago 
by Adler and Dothan~\cite{Adler:1966} and 
Wolfenstein~\cite{Wolfenstein:1970}.  
N. Kaiser used HBChPT to show that the next order corrections to $G_P$ are 
very small~\cite{Kaiser:2003}. 
One challenge exists:   
Can $G_P(q^2)$ 
be measured in some process in order to 
confirm this theoretical prediction? 

Two processes can determine $G_P$, OMC and RMC. 
The $\mu^- p$ capture rate has recently been measured at PSI by 
Andreev {\it et al.}~\cite{Andreev:2007}.   
Instead of the standard liquid  Hydrogen target they~\cite{Andreev:2007} 
used a gas target in order to minimize the  molecular 
complications in the capture process, see e.g. Refs.~\cite{Ando:2000,Bernard:2002}.
The initial results are  
\underline{consistent} with the ChPT prediction. 
Forthcoming final experimental results are expected at 1\% accuracy. 
The radiative muon capture has the advantage that  $q$ changes with the 
photon energy,  $E_\gamma$,  meaning RMC 
could determine $G_P(q^2)$ via the pion-pole dominance of  
Eq.(\ref{eq:RMC}).  
A TRIUMF team was able to measure the extremely low  RMC rate, 
d$\Gamma /{\rm d} E_\gamma$, for photon energies 
$E_\gamma > 60 $ MeV~\cite{Jonkmans:1996,Wright:1998}. 
It was a big surprise that the   
RMC experimental results disagreed with the HBChPT prediction. 

The advantage of the systematic ChPT expansion can be illustrated by 
the following order by order expression for the $\mu^-p$ 
spin-singlet capture rate 
taken from Ref.~\cite{Bernard:2001}
%The expansion converges rapidly for 
% 
\begin{eqnarray}
\Gamma &=&
\left(957 - \frac{245 {\rm GeV} }{m_N} + 
\left[ \frac{30.4 {\rm GeV}^2 }{m_N^2} - 43.17\right] 
\right) s^{-1} \nonumber 
\end{eqnarray} 
The {\it near cancellation} of the two terms in the square bracket,
 originating from 
``recoil" ($1/m_N^2$) 
and $q^2$ form-factor contributions, testifies to the value of 
the systematic perturbative expansion of HBChPT.  \\

{\it The radiative corrections to 
neutron $\beta$-decay and the CHOOZ process} are 
of critical importance since 
in the coming decade   the processes
 $n\to p+e^-+\bar{\nu}_e$ 
and  $\bar{\nu}_e + p\to e^+ + n$   
will be 
measured very precisely. 
The precise measurements of neutron $\beta$-decay aim at  
an accurate value for  $V_{ud}$.  
To extract $V_{ud}$ requires an updated understanding of the 
radiative corrections (RC). 
The second reaction, 
the CHOOZ process~\cite{chooz:2002}, will be used to determine neutrino 
oscillation parameters.
Why a new investigation of these  RC?  
A systematic reevaluation of RC~\cite{vogel:1984} to the
CHOOZ process is possible within HBChPT, which  
allows a model-independent, gauge-invariant 
evaluation of RC. 
The {\it short distance physics} is again well defined 
in the HBChPT lagrangian by the radiative LECs, 
which  are determined in, e.g.,  
neutron beta-decay RC evaluation~\cite{Ando:2004}. \\ 

%%%%%%%%%%%%%%%%%%%%%%%%%%%%%%%%%%%%%%%%%%%%%%%%%%%%%%%%%%%%%%%
%\subsubsection*{Two-nucleon processes} %%%%%%%%%%%%%%%%%%%%%%%
%%%%%%%%%%%%%%%%%%%%%%%%%%%%%%%%%%%%%%%%%%%%%%%%%%%%%%%%%%%%%%% 
{\it The two-nucleon processes} to be discussed are 
connected to fundamental astro-physical reactions; 
%
%\begin{itemize}
%\item 
muon capture on the deuteron: 
$\mu^-+d \to \nu_\mu+n+n$, 
%\item 
the charged- and neutral currents (CC and NC) of the 
Sudbury Neutrino Observatory (SNO) reactions: 
$\nu_e+d\to e^-+p+p \; \; $ and 
 $\; \; \nu_x+d\to \nu_x+p+n$, 
%\item
and the radiative pion capture on the deuteron:  
 $\pi^- + d \to \gamma + n + n$  or the crossing symmetric process 
$\gamma + d \to \pi^+ + n + n$. 
%\end{itemize} 
%
The ChPT evaluation of these reactions   
include one unknown axial {\it two-nucleon}  LEC,
$\hat{d}^R$, which  
%The  $\hat{d}^R$ 
also enters in the evaluations of the 
following few-nucleon reactions~\cite{Park:2003}; 
%\begin{itemize}
%\item
triton $\beta$-decay: $^3H\to ^3He + e^+ + \nu_e$, 
%\item
solar $pp$ fusion: $p+p\to d+e^++\nu_e$,  
%\item
the solar Hep process: $^3He+p\to ^4He+e^++\nu_e$, and the 
%\item
modern three-nucleon potential ($\hat{d}^R$ is related to $c_D$, one of the 
two unknown LEC parameters in the 
ChPT-derived three-nucleon potential~\cite{Nogga:2004}).
%\end{itemize}
The Hep process produces the highest 
energy solar neutrinos and has to be carefully evaluated~\cite{Kubodera:2004} 
in order to extract accurately 
the $^8$Be solar neutrino spectrum detected at, e.g., SuperKamiokande and SNO.
A precise evaluation of Hep is however difficult since 
leading contributions almost cancel as discussed in e.g.~\cite{Carlson:1991}.

Ideally the two-nucleon reactions should be evaluated using 
transition operators and nucleon wave functions obtained from ChPT. For 
pragmatic reasons however 
a  {\it hybrid} ChPT called $EFT^*$ has been used in 
the two- and more nucleon processes.  
In $EFT^*$  we use 
%\begin{itemize}
%\item
the one- and two-nucleon transition operators from ChPT, 
whereas   
%\item
the nuclear wave functions are evaluated using modern 
``high precision" $NN$ potentials $V_{NN}$, 
e.g., Argonne $V_{18}$, CD-Bonn, $V_{low-k}$, etc.
%\end{itemize}
%
%The reaction rates and cross-sections are all evaluated within 
%the same $ECT^*$ formalism. 
%
In  $EFT^*$ calculations a 
Gaussian cut-off $\Lambda_G$ was introduced in the 
nuclear wave functions in order 
to limit the contributions from the 
high momentum components of the wave functions 
generated by $V_{NN}$. 
%ChPT assumes that the high momentum components have been ``integrated out". 
%
These high momentum components in the nuclear wave functions 
generated by, e.g., the Argonne $V_{18}$ potential, go beyond   
the relevant limited momentum range of ChPT, 
$Q^2 \ll \Lambda_{ch}$. 
This Gaussian cut-off procedure is therefore in accordance 
with one of the principal assumptions of ChPT allowing 
only a limited low $Q^2$ range.
As a consequence however 
the axial two-nucleon LEC will $\hat{d}^R$ depend on $\Lambda_G$. 
The observables should be independent of this momentum cut-off, and we 
find that the measurable rates and cross-sections have  
less than 1\% variations  
for 500 MeV $ < \Lambda_G <$ 800 MeV.

Presently $\hat{d}^R$ is 
is determined from 
%a  \underline{three-nucleon} system: 
%\\ 
tritium $\beta$-decay rate. 
It is however desirable to avoid the complexity of a three-nucleon 
system in determining  the  \underline{two-nucleon} 
axial coupling  $\hat{d}^R$, so that two-nucleon 
processes can be calculated self-consistently within the 
framework of ChPT.  
Avoiding the inherent uncertainties of the three-nucleon system 
will also allow a more reliable evaluation of the 
uncertainties involved in two-nucleon reactions.
The rate of muon capture on a deuteron ($\mu^- d$) 
is being measured (2009-2011) at PSI by the 
MuSun collaboration with a projected  
error of 1.5\%~\cite{Kammel:2010}. 
We 
%\footnote{ A. G\aa rdestig, K. Kubodera and F. Myhrer} 
are presently re-evaluating our  
$\mu^- d$ ChPT calculation to match this expected experimental precision. 
Once the $\mu^- d$ capture rate is accurately measured, 
the following 
three reaction can be evaluated model independently with the same accuracy: 
(i) {\it the  
solar $pp$ fusion reaction}, the primary energy source in the sun, 
(ii)  
{\it the 
SNO neutrino-deuteron reactions}, 
which provided convincing evidence for 
neutrino oscillation,  
and 
 (iii) the reaction $\pi^- + d \to \gamma + n +n$~\cite{Gardestig:2006} or 
$\gamma + d \to \pi^+ + n + n$~\cite{Lensky:2005} which 
can be used to determine  
the neutron-neutron scattering length $a_{nn}$, 
see the review~\cite{Gardestig:2009} for a discussion. 
Furthermore, one of the LEC in the three-nucleon potential, $c_D$, which 
is an axial two-nucleon LEC, is 
determined once the value of $\hat{d}^R$ 
is fixed by the $\mu^- d$ capture reaction. 
In other words, only one unknown three-nucleon LEC, $c_E$, 
remains in the ChPT three-nucleon potential. 

The expected accurate measurements of the $\mu^- p$ and $\mu^- d$ 
capture rates will require a re-examination of the radiative corrections 
to these two processes. 
The estimated radiative corrections are larger than the expected 
experimental errors from the MuCap and MuSun 
collaborations and a renewed evaluation of the RC 
is in progress. 

{\it Supernova Explosion} 

Computer simulations of the supernova have not been very 
successful in generating the explosion. 
This is possibly due to the neutrino luminosity being too small. 
We have identified new processes which generate neutrinos 
in the proto-neutron star 
at the center of the supernova explosion. 
These reactions might affect 
the explosion-simulation due to  
an (estimated) increased 
in the neutrino flux~\cite{sato:2010}.

\subsection*{Conclusions}

The low-energy effective theory, ChPT, 
allows a systematic evaluation of electro-weak 
and strong interaction processes. 
%\item
HBChPT predicts accurately the analytic expression for $G_P(q^2)$. 
The predicted value for $G_P$ in the $\mu^- p$ process is  
confirmed by recent MuCap data~\cite{Andreev:2007}. 
The published  MuCap experimental $\mu^- p$ capture 
rate is also compatible with the HBChPT prediction. 
The advantage utilizing ChPT is that 
ChPT provides analytic expressions for 
both the $\mu^- p$ and $\mu^- d$ capture operators 
at each perturbative order,  
and 
ChPT permits us to make a reasonable estimate the theoretical uncertainty 
of the calculated observable.  
Once we have a 
measurable quantity evaluated at ``order" $\left(Q/\Lambda_{ch}\right)^v$, 
an estimated uncertainty is given by 
the magnitude of the next order contribution
$\left(Q/\Lambda_{ch}\right)^{v+1}$. 
%\item
Two-nucleons reactions (including the energy dependence of 
$n+p\to d+\gamma$ which is important in cosmology) are well described by 
 $EFT^*$ (one exception is the measured RMC rate versus the photon energy). 
The $\mu^- d$ capture process being measured by the MuSun collaboration at PSI
will allow a more accurate value for $\hat{d}_R$. 
This MuSun measurement will permit us to make  more 
accurate model-independent predictions for the solar $pp$ fusion and the 
$\nu d$ SNO reactions. 
However, improved radiative corrections are needed to ``compete" with the 
expected MuCap and MuSun $\mu^- $-capture data.
ChPT is ideally suited for an evaluation of these radiative corrections.

\begin{theacknowledgments}
This work was supported in part by the NSF grant PHY-0758114.
\end{theacknowledgments}

%%%%%%%%%%%%%%%%%%%%%%%%%%%%%%%%%%%%%%%%%%%%%%%%%%%%%%%%%%%%%%

\end{document}